\newcommand{\beq}{  \begin{eqnarray}}
\newcommand{\eeq}{  \end{eqnarray}}

\documentclass[prl,twocolumn,showpacs,preprintnumbers,amsmath,amssymb]{revtex4-1}
\usepackage{subfig}
\usepackage{graphicx}
\usepackage{dcolumn}
\usepackage{bm}

\begin{document}
    	
\title{Geometric Phase Optics}
    
\author{B. Zygelman}
\email{bernard@physics.unlv.edu}
\affiliation{%
Department of Physics and Astronomy, University of Nevada, Las Vegas, Las Vegas NV 89154
}%

\date{\today}
\begin{abstract}
 We introduce, and propagate wave-packet solutions of, a single qubit system in which geometric gauge forces and phases
 emerge. We investigate under what conditions non-trivial gauge phenomena arise, and demonstrate how symmetry breaking
 is an essential ingredient for realization of the former. We illustrate how a ``magnetic''-lens, for neutral atoms, can be constructed
 and find application in the manipulation and interferometry of cold atoms.
\end{abstract}

\maketitle
\section{Introduction}
Gauge symmetry is at the core of our current understanding of how the fundamental constituents of
matter interact.  With the discovery of the geometric phase\cite{shap89} diverse systems, ranging 
atomic, molecular, optical, condensed matter and nuclear physics, have been identified in which gauge phenomena, in addition to those
arising from fundamental gauge fields,  emerge.  More recently, researchers\cite{lin09a} have, via the application of laser fields on cold atoms,
engineered Hamiltonians that lead to effective ``magnetic''-like forces on the atoms.
This advance has great potential in the control and manipulation of quantum matter\cite{dal10b}.  
In particular, its application promises the capability to create 
ensembles of neutral atoms that exhibit exotic quantum Hall-like behavior\cite{diaz11,spiel11}.
 
In this Letter, we introduce a single qubit model that possesses non-trivial gauge behavior and whose laboratory realization
may offer novel routes to the quantum control, and interferometry, of cold atoms. 
We present results of time-dependent calculations for wave-packet 
propagation to illustrate how, and under what conditions, geometric gauge forces manifest.  We show how the proposed system mimics
that of a charged particle scattered by a ferromagnetic medium. We illustrate how an effective ``magnetic'' lens can be engineered and propose
possible applications.   
\section{Theory}
Consider the Hamiltonian
\beq
H= - \, \frac{\hbar^{2}}{2m} \, {\bm \nabla}^{2} + H_{ad}({\bm R}) 
\label{1.1}
\eeq
where $H_{ad}({\bm R})$, the adiabatic Hamiltonian describing a qubit, is parameterized by the quantum
variable ${\bm R}$, and which can be  expressed in the form
\beq
H_{ad} = U({\bm R}) H_{BO} U^{\dag}({\bm R}).
\label{1.1b}
\eeq
A detailed, time-independent, description of such systems has been outlined in Ref.  \cite{zyg12}, but here we exploit 
time-dependent methods to enhance and generalize the conclusions of that paper.  Laboratory realizations
of adiabatic Hamiltonians discussed in this Letter could be achieved using the techniques discussed in Refs. \cite{dal10b,diaz11}.
Because $U$ is a unitary operator the 
eigenvalues of $H_{ad}$ are solely determined by $H_{BO}$  that, in this study, we require to be non-degenerate. 
We take the eigenstates of $H_{BO}$ as our basis and  set 
$ H_{BO}=\Delta \, \sigma_{3} $ where $ \Delta >0 $ is a constant and $\sigma_{3}$ is
the diagonal Pauli matrix. We choose\cite{zyg12} for $U$,
\beq
\exp(-i \, \sigma_{3} \frac{\Phi}{2} \, y) \exp(-i \sigma_{2} \, \Omega(x) )  \exp(i \, \sigma_{3} \frac{\Phi}{2} \, y)
\label{1.2}
\eeq
where $\sigma_{i}$ are the Pauli matrices, $\Phi$ is a parameter, $ {\bm R}= (x,y) $ are 2D Cartesian coordinates, and 
\beq
\Omega(x) = \frac{\pi}{4} \, \Bigl ( 1+ \tanh(\beta x) \Bigr ).
\label{1.3}
\eeq
We seek solutions of $ i \,\hbar \, \partial_{t} \psi({\bm R}) = H \psi({\bm R}) $ which, expressed in this
basis, take the form of the coupled equations 
\beq
&& i \hbar \, \frac{\partial f}{\partial t} = -\frac{\hbar^{2}}{2 m} {\bm \nabla}^{2} f + V \, f + V_{12} g \nonumber \\
&& i \hbar \, \frac{\partial g}{\partial t}= -\frac{\hbar^{2}}{2 m} {\bm \nabla}^{2} g + V^{*}_{12} g - V \, f
\label{1.4}
\eeq
where 
\beq
&& \psi \equiv \left ( \begin{array}{c} f \\ g \end{array} \right ) \nonumber \\
&& V= \Delta \, \cos(2 \Omega(x)) \nonumber \\
&& V_{12}= \exp(- \, i \, \Phi \, y) \, \Delta \, \sin(2 \Omega(x)). 
\label{1.5}
\eeq
They may be solved using the split-operator method\cite{fleck88}, and a wave packet at
$ t=t_{0}$, can be propagated to $ t=t_{0}+\delta t$, for a small time increment $\delta t$.
Introducing the dimensionless units $ \tau = \frac{L^{2}}{2 m \hbar}\, t$, $ (\xi,\eta)=(x/L,y/L) $, where $L$ is
an arbitrary length scale, we obtain
\beq
\psi(\tau+\delta \tau)=U_{KE} U_{V} U_{KE} \, \psi(\tau) 
\label{1.6}
\eeq
where $U_{KE}$ is a diagonal matrix operator whose elements are,
\beq
&& \exp(i \, \frac{\delta \tau}{2}  \Bigl ( \frac{\partial^{2}}{\partial \xi^{2}} +  \frac{\partial^{2}}{\partial \eta^{2}} \Bigr ) ) \nonumber
\eeq
and
\begin{widetext}
\beq
U_{V}= & \left ( \begin{array}{cc} \cos(\Delta  \,  \delta \tau) - 
i \cos(2 \Omega(\xi)) \sin(\Delta  \, \delta \tau)
  & -i \, \exp(- i \Phi \, \eta) \sin(2 \Omega(\xi)) \sin(\Delta   \, \delta \tau) \\
 -i \, \exp(i \Phi \, \eta) \sin(2 \Omega(\xi)) \sin(\Delta  \, \delta \tau)
& \cos(\Delta  \, \tau) + i \cos(2 \Omega(\xi)) \sin(\Delta  \,  
\delta \tau) \end{array} \right ). \nonumber
\eeq
\end{widetext}

\begin{figure}[ht]
\centering
\includegraphics[width=0.9\linewidth]{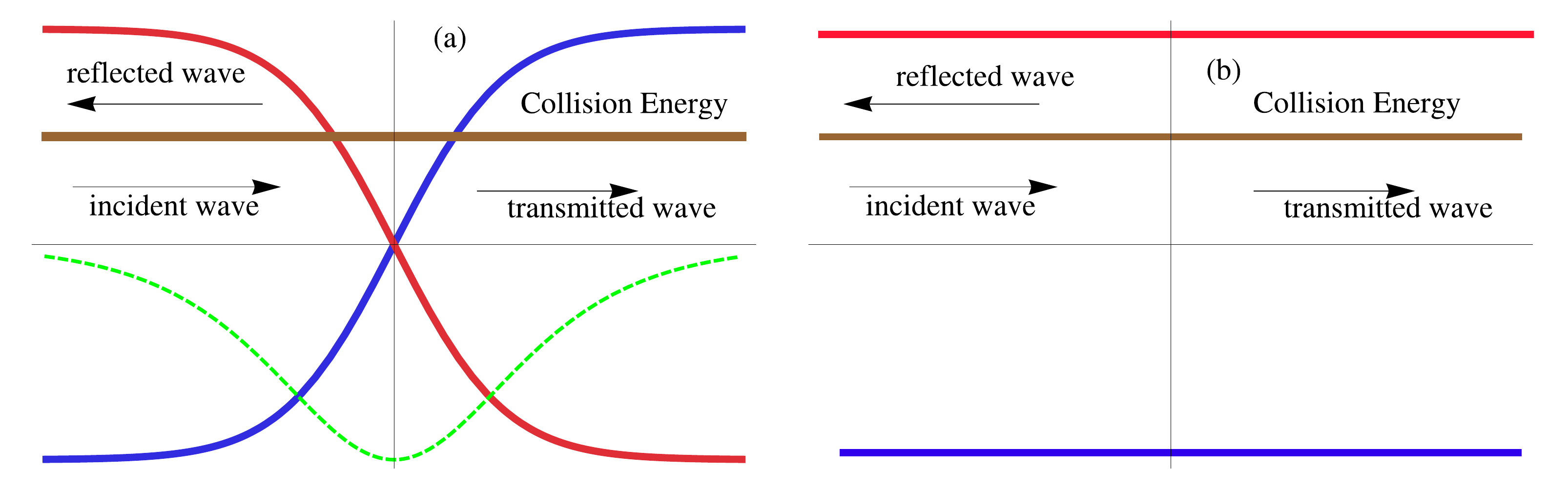}
\caption{\label{fig:fig1}(a) An incident wave in the open channel whose potential energy is 
given by the solid blue line. For $\xi>0$ the transmitted wave is propagated on the potential surface given by the red solid line. The brown line represent the total
collision energy of the system and the green dashed line represents the off-diagonal coupling between the two potential surfaces.
(b) The same system now illustrated in the adiabatic picture (gauge). The blue line is the BO energy for the open channel and the red line represents
the BO energy for the closed channel.}
\end{figure}

\begin{figure*}[ht]
\centering
\includegraphics[width=0.5\textwidth]{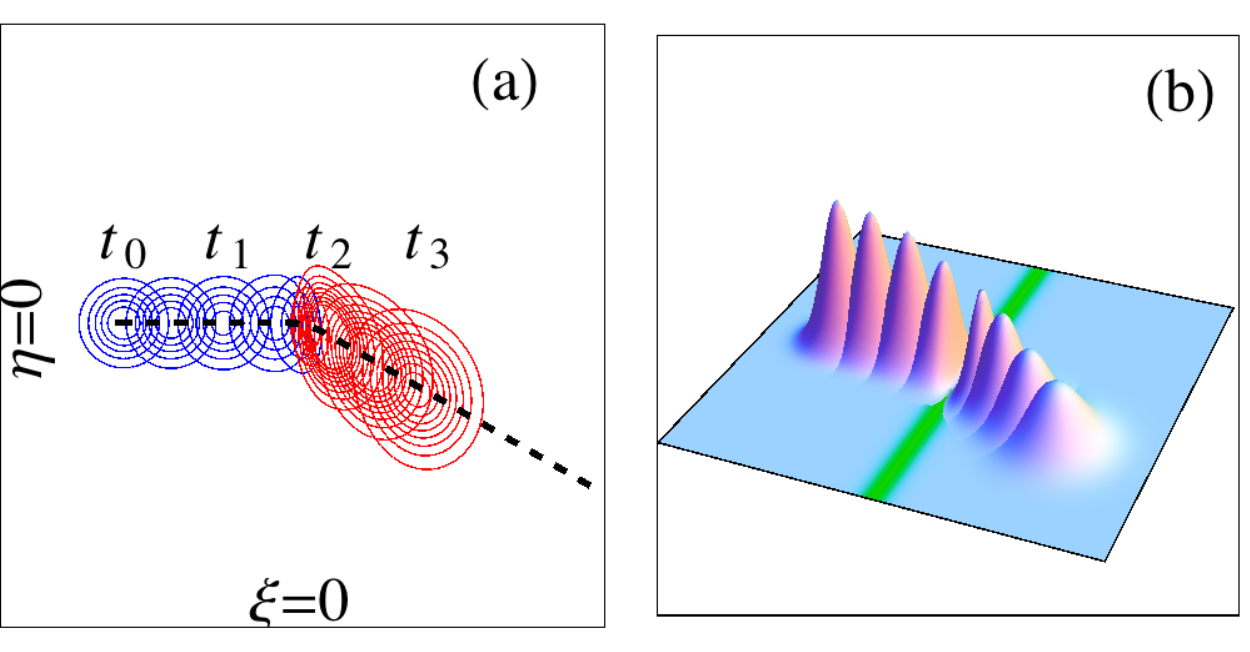}
\caption{\label{fig:fig2} (a) Blue contours represent the initial $g$ component of the wave packet probability distribution.
 At time $ t_{2}$ the packet executes a transition into the $f$ channel, shown by the red contours, which is energetically open in the region $ \xi >0$. 
Subsequent to the transition the packet evolves as a free particle but with a pronounced swerve in its velocity. 
(b) 3D plot of the adiabatic gauge probability distribution  
$|\tilde g(t)|^{2}$. In both figures the deflection angle has the value $\tan \theta \approx 0.59 $. 
 In these calculations we took $k=12,\Phi=6,\Delta=200,\beta=2$ and $ {\tau}_{0}=0, {\tau}_{1} =20,{\tau}_{2}=30,{\tau}_{3}=40$. $\xi,\eta$ range between $\mp 2\pi$.   }
\end{figure*}

Expressed in these units $ \Delta $ is a dimensionless parameter as are $\Phi,\beta$.
Figure (\ref{fig:fig1}) provides a schematic description of the dynamics generated by Hamiltonian (\ref{1.1}).
In Figure (\ref{fig:fig2}a) we provide a time series contour plot of the probability densities $|f|^{2}$ and $|g|^{2}$.
At $\tau=0$ we place a Gaussian wave-packet $\psi_{0}=\left ( \begin{array}{c} 0 \\ g_{0} \end{array} \right )$  centered 
$\xi=-4.0, \, \eta=0$ with an initial velocity directed along
the positive $\xi$ axis. In this region ($ \Omega \rightarrow 0 $)
\beq
 H_{ad} = \left ( \begin{array}{cc} \Delta & 0 \\   0 & -\Delta  \end{array} \right ) 
\nonumber
\eeq
and the wave packet evolves as that of a free particle until it reaches the interaction region  $ \xi \approx 0$. 
The wave-packet is illustrated by the blue contours in Figure (\ref{fig:fig2}a).
The initial kinetic energy of the packet was chosen so penetration of the potential barrier, illustrated in Figure (\ref{fig:fig1}), 
is prevented. However,  the packet can execute a transition into the open  channel across the barrier. In other words, a transition
from the $ g $ to $ f$ channel  occurs in the region $\xi \approx 0$. This is illustrated in Figure (\ref{fig:fig2}a) by the red
contours that represent the wave-packet, probability, contours in the $f$ channel. In addition to distortion and spreading of the packet there
is a noticeable swerve in its velocity as it emerges from the interaction region.

We define the adiabatic amplitudes,
\beq
{\tilde f} = f \, \cos\Omega(\xi) + \exp(-i \Phi \eta) \sin\Omega(\xi) g \nonumber \\
{\tilde g} = g \, \cos\Omega(\xi) - \exp(i \Phi \eta) \sin\Omega(\xi) f
\label{1.7}
\eeq
and
${\tilde \psi} = \left ( \begin{array}{c} {\tilde f} \\ {\tilde g} \end{array} \right ) $ obeys the following
equation\cite{zyg12} 
\beq
i \hbar \frac{\partial {\tilde \psi }}{\partial t} =
-\frac{\hbar^{2}}{2 m} \Bigl ( {\bm \nabla }- i {\bm A}  \Bigr )^{2} {\tilde \psi} + H_{BO} {\tilde \psi},   
\label{1.8}
\eeq
where $ {\bm A}$ is a non-Abelian, pure, gauge potential. In the region $ \xi \rightarrow -\infty$, $ \Omega \rightarrow 0$
and ${\tilde g}  \rightarrow g$. Likewise as $ \xi \rightarrow \infty$, $ \Omega \rightarrow \pi/2 $ and 
$ {\tilde g} \rightarrow f $. This behavior is illustrated in Figure (\ref{fig:fig2}b) where we present a 3D plot
for the evolution of $ |{\tilde g}|^{2}$. In the adiabatic picture the open channel amplitude ${\tilde g}$ evolves
in a  constant adiabatic potential $ -\Delta $ shown in Figure (\ref{fig:fig1}b).  As long as the collision energy is below
the threshold for excitation into the upper adiabatic, or closed, channel the system evolves on a single adiabatic
surface. Under such conditions the Born-Oppenheimer (BO) approximation to the solutions of Eq. (\ref{1.8}) is
appropriate. In this approximation, the projection operator $ P {\tilde \psi} = {\tilde g}$, is applied on 
 Eq. (\ref{1.8}) to get
 \beq
 i \hbar \frac{\partial {\tilde g }}{\partial t} = -\frac{\hbar^{2}}{2 m}  \Bigl ( {\bm \nabla }- 
i {\tilde {\bm  A} }  \Bigr )^{2} {\tilde g} -\Bigl  (\Delta- \frac{\hbar^{2} \, b(x)}{2 m}  \Bigr )  {\tilde g}  
\label{1.9}
\eeq
where $  {\tilde {\bm A} } = P \,  {\bm A} \, P  $ is an Abelian gauge potential with non-vanishing curl and $b(x)$ is an induced scalar potential\cite{zyg12}. 
It leads to an effective magnetic induction 
\beq
|{\bm B}(\xi)|= \frac{\pi}{4} \, \beta \, \Phi \, \text{sech}^{2}(\beta \xi) \cos( \frac{\pi}{2} \tanh(\beta \xi)) 
\label{1.9a}
\eeq
and mimics
that incurred on a charged particle that is scattered by a ferromagnetic medium.
 The magnetic induction is normal to the plane of the page, 
 and is illustrated by the green shaded area in Figure (\ref{fig:fig2}b).
In Figure (\ref{fig:fig2}a) we also plot, shown by the dashed line, the 
trajectory for the solution of the classical equations of motion, subjected to a Lorentz force $ {\bm v} \times {\bm B}$, where
$|{\bm B}|$ is given by Eq. (\ref{1.9a}).
Comparison of the classical path and that traced by 
the centers of the wave-packets shows good agreement.
\begin{figure}[ht]
\centering
\includegraphics[width=0.9\linewidth]{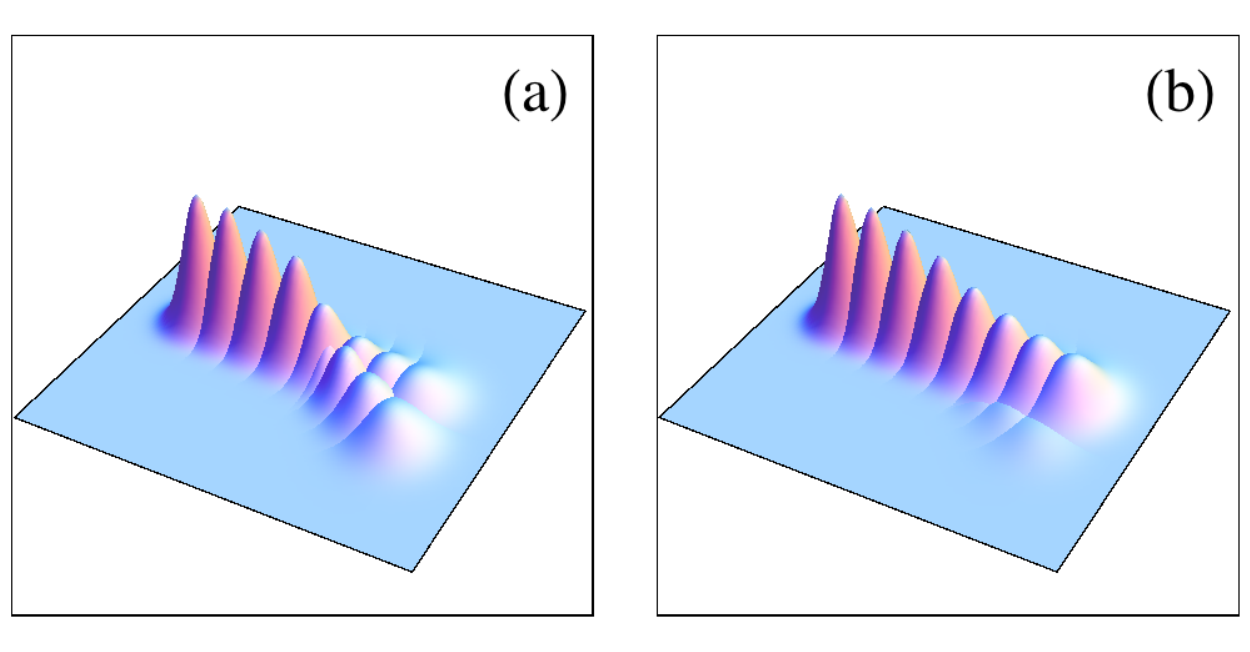}
\caption{\label{fig:fig3} Plot of probability densities $|g|^{2},|f|^{2}$ when both channels are open. The threshold energy $k_{t}^{2}=2 \Delta$. 
(a) $ k^{2} = 3 k_{t}^{2}$. (b) $k^{2}= 6 k_{t}^{2}$.} 
\end{figure}
The deflection angle 
suffered by a charged particle that is normally incident on a ferromagnetic slab, of finite width, 
with constant magnetic induction ${\bm B}$ directed
perpendicular to the plane of the page is\cite{zyg12}, 
\beq
\tan \theta = \frac{ |{\Phi}| }{\sqrt{k^{2} - \Phi^{2}}} 
\label{1.11}
\eeq
where $ \Phi = \int_{-\infty}^{\infty} d \xi \, B(\xi) $ is a flux density, and $k$ the incident wave number.

In Table \ref{tab:Table1} we tabulate values of the deflection angles $\theta$, obtained by calculating the expectation values
$<\xi(t)>$, $<\eta(t)>$ for various values of incident, adiabatic, packet wave numbers
$ k$ and $\Phi$. In that table we show the dependence of $\theta$ on the choice of the energy gap parameter $\Delta$.
At lower collision energies, so that $ k^{2} << 2 \Delta $, we find that Eq. (\ref{1.11})  provides a good approximation for $\theta$.
As the energy gap $2 \Delta$ is decreased, for a fixed value of $k$, Eq. (\ref{1.11}) is less accurate.
However, even at threshold $ k^{2} \approx 2 \Delta$ there is still fairly good agreement between the calculated value and that predicted
by solutions of Eq. (\ref{1.9}). When $ k^{2} > 2 \Delta$ the excited adiabatic state
is open and transitions from the adiabatic channel labeled $\tilde g$ into $\tilde f$ is energetically allowed. In Figure (\ref{fig:fig3}a) we
illustrate the evolution of the amplitudes $ |g(t)|^{2} $ and $ |f(t)|^{2} $ for the collision energy where $ 2 \Delta/k^{2} = 1/3 $. The
incident packet, in the $g$ channel, bifurcates when it reaches the interaction region. Because there is sufficient kinetic energy, the
remainder of the initial packet proceeds along the path $\eta=0$ in the region $ \xi >0$. However, a fraction of that packet makes
a transition into the $f$ channel, and our calculations show that the angle of the swerve illustrated in that figure is in harmony
with that obtained at the lower collision energies tabulated in Table \ref{tab:Table1}. Therefore there is a state-dependent spatial
segregation of the initial beam, a hallmark of quantum control. In panel (b) of that figure we plot
these probabilities for energies $ 2 \Delta/k^{2} = 1/6$ and now find a small, barely noticeable, remnant of the packet in the $f$ channel.
In the limit $ k \rightarrow \infty$  (or $\Delta \rightarrow 0$) Eq. (\ref{1.4}) allow analytic solutions and $g(t)$ simply
evolves as that of a free particle. According to definition Eq. (\ref{1.7} ) the initial, adiabatic gauge, packet $ {\tilde g}(t) $ makes 
a transition into the $ {\tilde f}$ channel in the region $\xi > 0$. This ''transition'' is induced by the off-diagonal
gauge couplings in Eq. (\ref{1.8}). The  "transition" is simply an artifact of the adiabatic gauge (i.e. different definitions
for the scattering basis in the two asymptotic regions, $\xi<0$, $\xi>0$) and does constitute a ``real'' physical transition. 
\begin{table}
\caption{\label{tab:Table1} Calculated values for the deflection angle $\theta$ are tabulated
in the third column. The fourth column gives the values obtained using Eq. (\ref{1.11})}.
\begin{ruledtabular}
\begin{tabular}{llll}
$2 \Delta/k^{2} $  &  $\Phi/k $  &  $\tan \theta $ &  $\tan \theta_{c} $   \\
\hline
$25/9$ & $1/2$ & $0.587$  &  $0.577  $    \\
$ 25/9$ & $1/4$ & $0.270$ & $0.258 $ \\ 
$ 25/9$ &  $1/12$ & $0.088  $ & $ 0.084 $ \\
\hline
$1$ & $1/2$ & $0.63$  &  $0.577  $    \\
$1$ & $1/4$ & $0.269$ & $0.258 $ \\ 
$1$ &  $1/12$ & $0.088  $ & $ 0.084 $ \\
\end{tabular}
\end{ruledtabular}
\end{table}
In order to better understand the behavior illustrated above we re-express the unitary operator $U$ that defines
the adiabatic Hamiltonian. In Ref. \cite{zyg12} we argued that $U$ can always be written in the form
\beq
U({\bm R})= {\cal P} \exp(-i \int^{{\bm R}}_{\cal C} {d \bm R'} \cdot {\bm A} )
\label{1.12}
\eeq
where ${\cal P}$ is a path ordering operator, and $ {\bm A}$ is a non-Abelian gauge potential. 
$U$ must be well-defined (i.e. not multi-valued) for all ${\bm R}$ and therefore Eq. (\ref{1.12}) must be
 independent of the path ${\cal C}$, or $ {\cal P} \exp(-i \oint {d \bm R'} \cdot {\bm A} ) =1 $. 
 Gauge potentials that satisfy this condition are sometimes called
a pure gauge and typically have vanishing curvature everywhere.  Because of relation (\ref{1.12}) we conclude that ${\bm A}$ is
encoded in the definition of $H_{ad}$ and since $ [H_{BO},{\bm A}] \neq 0$ gauge symmetry is explicitly broken by $H_{BO}$.
Though ${\bm A}$ is trivial, in the sense of it being a pure gauge, 
quantum evolution selects and is sensitive to the projected ${\tilde {\bm A}} = P {\bm A} P$ {\bf \it non-trivial} connection.  In the
adiabatic picture the gauge potentials are explicit, being minimally coupled to the amplitudes.  As $ k \rightarrow \infty$, or $ \Delta \rightarrow 0$,
their presence simply contributes to a multichannel, or non-Abelian, phase in the adiabatic amplitudes that has
 no physical import. In contrast,
at lower energies the system behaves as if it has acquired a non-integrable phase factor. The effect is most pronounced when the excited
adiabatic state is closed.
\section{Applications}
\begin{figure}[ht]
\centering
\includegraphics[width=0.9\linewidth]{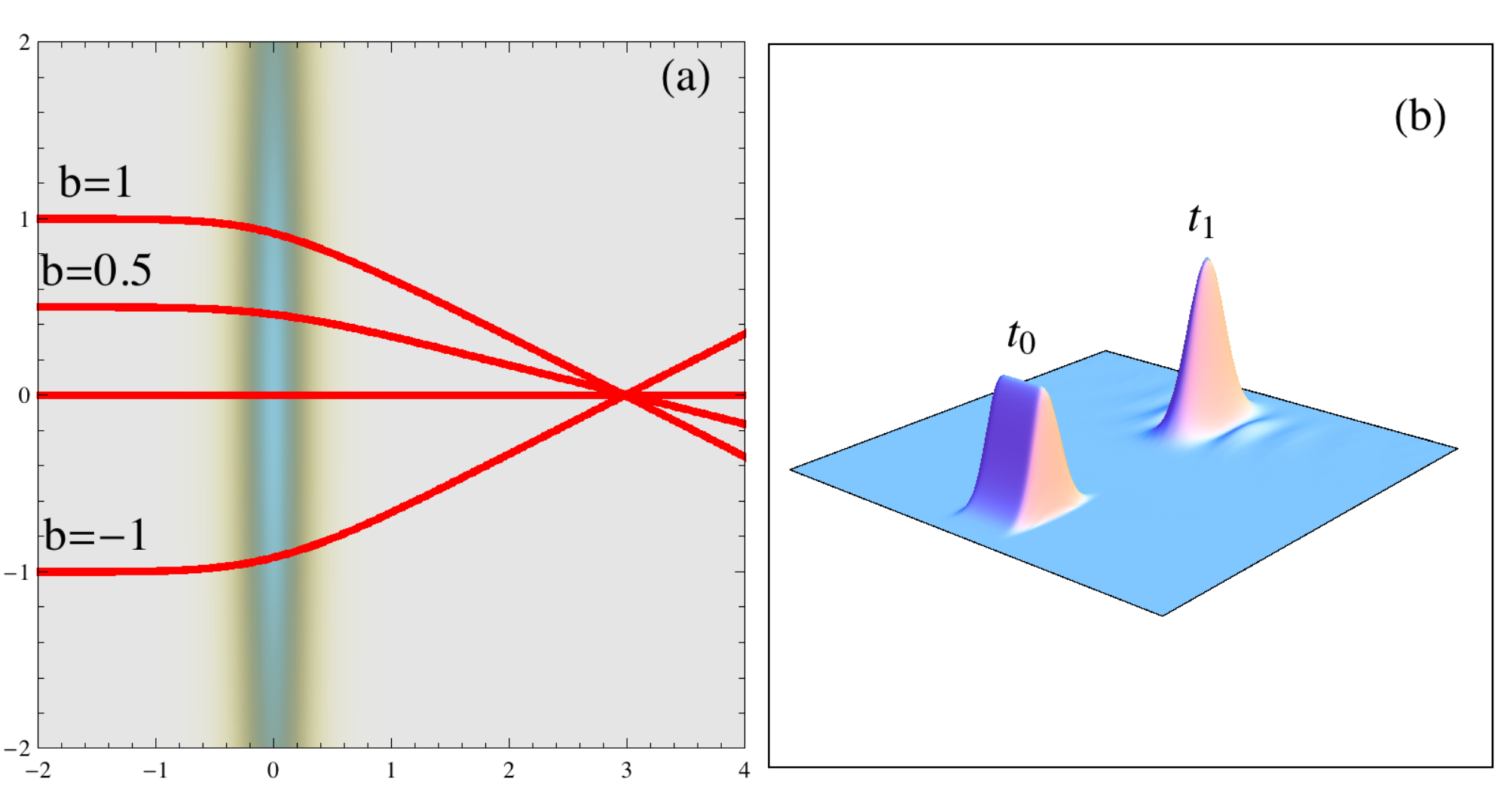}
\caption{\label{fig:fig4} (a) Trajectories of wave-packet expectation values for various values of impact parameter. The shaded region
is a density plot of Eq. (\ref{2.2}) for the ``magnetic'' induction. (b) A wave packet slab having width $d=2$ along the $\eta$ axis at ${\tau}_{0}=0$
is propagated to the position shown at ${\tau}_{1}=30$.}
\end{figure}

In the discussion up to this point, we have used 
time-dependent methods to validate and extend the conclusions given in Ref. \cite{zyg12} in
which time-independent methods allow exact scattering solutions for Hamiltonian (\ref{1.1}). However, time
dependent methods can be exploited for more complex scattering scenarios. Following an analysis similar to that in which
Eq. (\ref{1.11}) was derived, we now posit the following form for the parameter 
\beq
\Phi(\eta) = \frac{\eta \, k}{\sqrt{\eta^{2} + 4 \gamma f^{2}}}
\label{2.1}
\eeq
where $\gamma$ is, in general, a complicated function of $\beta, k, \Delta$. Here 
we set it to have the constant  value $\gamma =1$. Using Eq. (\ref{2.1}) we propagate wave packets for various
values of impact parameter. In Figure (\ref{fig:fig4} a) we plot trajectories of the total expectation values
$ <\psi(t)|\xi|\psi(t)>, <\psi(t)|\eta|\psi(t)>$ for the various impact parameters b. At each impact
parameter we choose identical wave-packet widths and set $k=12, \Delta=400$. The trajectories shown
in that figure, by the solid red lines, demonstrate that the paths converge to a common focal point given by $f=3$. 
This result is gauge invariant, i.e. it can
be obtained using amplitudes obtained in either diabatic or adiabatic gauges. However, 
the adiabatic picture provides a transparent physical description. For, in it, the system is accurately described by
Eq. (\ref{1.9}). That description includes the emergence of an effective magnetic induction ${\bm B}={\nabla} \times {\tilde {\bm A}}$ whose
magnitude is
\beq
\frac{\pi  \beta  k \eta \left(8 f^2+\eta^2\right) \text{sech}^2(\beta  \xi) \cos
   \left(\frac{1}{2} \pi  \tanh (\beta  \xi )\right)}{4 \left(4 f^2+\eta^2\right)^{3/2}}
\label{2.2}
\eeq
and is normal to the plane of the page. In Figure (\ref{fig:fig4}b) we illustrate the propagation of a coherent wave packet slab of finite width
along the $\eta$ direction. After its passage through the ``magnetic'' lens at $\xi \approx 0$,  its shape is significantly distorted.
 At $\tau_{1}=30$, where the packet describes free particle evolution, it assumes the shape of a ``shark-fin'' as
shown in that figure. The width, along the $\eta$ direction, is significantly reduced from its value at $\tau_{0}$. A dramatic consequence
of the proposed ``magnetic'' lensing effect. Such a lens, if realized, could find application as an ``optical'' component in an atom laser.
In addition, consider two localized but coherent packets spatially separated at $t=0$.  After passing through the lens they meet and interfere.
Because of different geometric phase histories the interference pattern depends on the ``magnetic'' flux enclosed
by the paths. One can therefore anticipate its application as a novel expression
of atom interferometry. 

\bibliographystyle{apsrev4-1}

\end{document}